\documentclass[preprint,showpacs,preprintnumbers,amsmath,amssymb]{revtex4}

\usepackage{graphicx}
\usepackage{dcolumn}
\usepackage{bm}
\usepackage[german,american]{babel}

\begin{document}

\title{Fitness, chance, and myths:
an objective view on soccer results}

\author{Andreas Heuer}
\author{Oliver Rubner}

\affiliation{University of M\"unster, Institute of Physical Chemistry,
Corrensstr. 30, D-48149 M\"unster, Germany}

\begin{abstract}

We analyze the time series of soccer matches  in a model-free way
using data for the German soccer league (Bundesliga). We argue
that the goal difference is a better measure for the overall
fitness of a team than the number of points. It is shown that the
time evolution of the table during a season can be interpreted as
a random walk with an underlying constant drift. Variations of the
overall fitness mainly occur during the summer break but not
during a season. The fitness correlation shows a long-time decay
on the scale of a quarter century. Some typical soccer myths are
analyzed in detail. It is shown that losing but no winning streaks
exist. For this analysis ideas from multidimensional NMR
experiments have been borrowed. Furthermore, beyond the general
home advantage there is no statistically relevant indication of a
team-specific home fitness. Based on these insights a
 framework for a statistical
characterization of the results of a soccer league is introduced
and some general consequences for the prediction of soccer results
are formulated.

\end{abstract}

 \maketitle

\section{Introduction}

In recent years physicists have started to investigate time
series, resulting from successive matches in sports leagues. In
this context several basic questions can be asked. Is the champion
always the best team? \cite{ben1,ben2,buch} How many matches have
to be played in a league so that (nearly) always the best team
becomes the champion? \cite{ben1,ben2} Does the distribution of
goals follow a Poisson distribution and what are possible
interpretations of the observed deviations? \cite{tolan,janke}. In
those studies it has been attempted to have a simplified view on
complex processes such as soccer matches in order to extract some
basic features like, e.g., scaling laws. Some empirical
observations such as fat tails in the goal distributions can be
related to other fields such as finance markets \cite{Stanley} and
have been described, e.g., by the Zipf-Mandelbrot law
\cite{Malacarne}. Actually, also in more general context the
analysis of sports events, e.g. under the aspect of extreme value
statistics, has successfully entered the domain of physicists
activities \cite{Suter}.

A more specific view has been attempted in detailed studies of the
course of a soccer season. In one type of models; see e.g. Refs.
\cite{Lee97,Dixon97,Dixon98,Rue00}, one introduces different
parameters to characterize a team (e.g. offensive fitness) which
can  be obtained via Monte-Carlo techniques. These parameters
are then estimated based on a Poisson assumption about the number of
goals of both teams. Within these models, which were mainly applied
to the English Premier league, some temporal weighting factors were
included to take into account possible time variations of the
different team parameters. These models are aimed to
make predictions for the goals in individual matches. In
\cite{Rue00} it is reported that based on a complex fitting
procedure the time scale of memory loss with respect to the
different variables is as short as 100 days.

 A second type of model assumes just one fitness parameter for each
team and the outcome (home win, draw, away win) is then predicted
after comparing the difference of the team fitness parameters with
some fixed parameters \cite{Koning00}. The model parameters are then
estimated based on the results of the whole season. Here, no
temporal evolution of the team parameter is involved. This very
simple model has been used in \cite{Dobson03} to check whether the
outcome of one match influences the outcome of the successive match.
Of course, this type of results is only relevant if the used model
indeed reflects the key ingredients of the real soccer events in a
correct way. It has been also attempted to analyse individual soccer
matches on a very detailed level, e.g., to estimate the effect of
tactical changes \cite{Hirotsu}

The approach, taken in this work, is somewhat different. Before
devising appropriate models, which will be done in subsequent
work, we first  attempt to use a model-free approach to learn
about some of the underlying statistical features of German soccer
(1. Bundesliga). However, the methods are general enough so that
they can be easily adapted to different soccer leagues or even
different types of sports. The analysis is exclusively based on
the knowledge of the final results of the individual matches.
Since much of the earlier work in this
field originates from groups with a statistics or economy background,
there is some room for the application of
complimentary concepts, more common in the physics community. Examples are
finite-size scaling, the analysis of 2-time correlation functions
or the use of more complex correlation functions to unravel the
properties of subensembles, as used, e.g., in previous 4D NMR
experiments \cite{Klaus,Wilhelm,epl}.

Four key goals are followed in this work. First, we ask about
appropriate observables to characterize the overall fitness of a
team. Second, using this observable we analyze the temporal
evolution of the fitness on different time scales. Third, we
quantify statistical and systematic features for the
interpretation of a league table and derive some general
properties of prediction procedures. Forth, we clarify the
validity of some soccer myths which are often used in the typical
soccer language, including serious newspapers, but never have been
fully checked about their objective validity. Does something like a
winning or losing streak exist? Do some teams have a specific home
fitness during one season?

The paper is organized as follows. In Sect.II we briefly outline
our data basis. The discussion of the different possible measures
of the overall fitness is found in Sect.III. In the next step the
temporal evolution of the fitness is analyzed (Sect.IV). In Sect.V
it is shown how the systematic differences in the team fitness can
be separated from the statistical effects of soccer matches and
how a general statistical characterization can be performed. In
Sect.VI we present a detailed discussion of some soccer myths.
Finally,  in Sect.VII we end with a discussion and a summary. In
two appendices more detailed results about a few aspects of our
analysis are presented.

\section{Data basis}

We have taken the results of the German Bundesliga from
http://www.bundesliga-statistik.de. For technical reasons we have
excluded the seasons 1963/64, 1964/65 and 1991/92 because these
were the seasons where the league contained more or less than 18
teams. Every team plays against any other team twice the season,
once at home and once away. If not mentioned otherwise we have
used the results starting from the season 1987/88. The reason is
that in earlier years the number of goals per season was somewhat
larger, resulting in slightly different statistical properties.

\section{Using goals or points to measure the team fitness?}

\subsection{General problem}

Naturally, a strict characterization of the team fitness is not
possible because human behavior is involved in a complex manner. A
soccer team tries to win as many matches as possible during a
season. Of course, teams with a better fitness will be more
successful in this endeavor. As a consequence the number of points
$P$ or the goal difference $\Delta G$ can be regarded as a measure
for the fitness. In what follows all observables are defined as
the average value per match.

In Sect.IV it is shown
that apart from fluctuations the team fitness remains constant during a season. Thus, in a
hypothetical season where teams play infinitely often against each other and thus statistical
effects are averaged out the values of $P$ indeed allow a strict sorting of the quality of the
teams. Thus, $P$ is a well-defined fitness measure for the team fitness during a season. Naturally,
the same holds for $\Delta G$ if the final ranking would be related to the goal difference. Since in reality the
champion is determined from the number of points one might tend to favor $P$ to characterize the team fitness.
In any event, one would expect that the rankings with respect to $\Delta G$ or $P$ are identical in this hypothetical limit.

Evidently, in a match the number of goals scored or conceded by a team is
governed by many unforeseen effects. This is one of the
reasons why soccer is so popular. As a consequence, the empirical values of
$P$ or $\Delta G$ obtained, e.g., after a full season will deviate from the limiting values due to the residual fluctuations.
This suggests a relevant criterion to distinguish between different observables. Which observable displays a minimum sensitivity
on statistical effects? As will be shown below, this  criterion favors the use of $\Delta G$.

\subsection{Distribution of $\Delta G$}

\begin{figure}
\includegraphics[width=7cm]{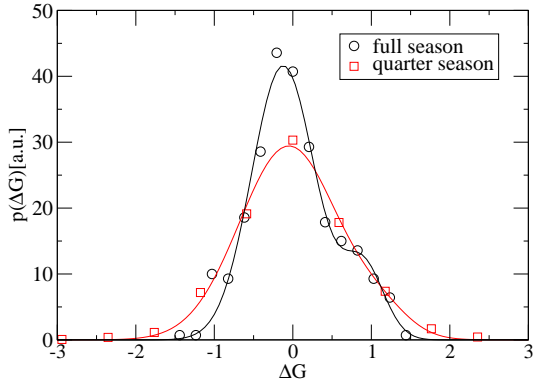}
\caption{\label{deltag_dist} The distribution of $\Delta G$ after
one quarter of the season and after a full season.  Included is a
fit with two Gaussian functions for both distributions.  For the
full-season distribution the intensity ratio of both Gaussian
curves is approx. 1:6. The correlation coefficient for the latter is 0.985.  }
\end{figure}

In Fig.\ref{deltag_dist} we display the distribution of
$\Delta G$ after one quarter of a season (thereby averaging over all
quarters) and at the end of the season. The first case corresponds
to $N=9$ (first and third quarter) or $8$ (second and fourth
quarter), the second case to $N=34$. Here $N$ denotes the number of
subsequent matches, included in the determination of $\Delta G$.

Both distributions can be described as a Gaussian plus an
additional wing at large $\Delta G$. Fitting each curve by a sum
of two Gaussians, the amplitude ratio for the full-season
distribution implies that there are on average 2-3 teams with an exceptional good fitness.

Note that the distribution of $\Delta G$ is significantly narrower
for larger $N$ and also for $N=34$ one expects some finite statistical
contribution to the width of the distribution.  Qualitatively, this reflects the statistical
nature of individual soccer matches. Naturally, the statistical
contribution becomes less relevant when averaging over more
matches. This averaging effect will be quantified in Sect.V.

\subsection{Correlation analysis}

\begin{figure}
\includegraphics[width=7cm]{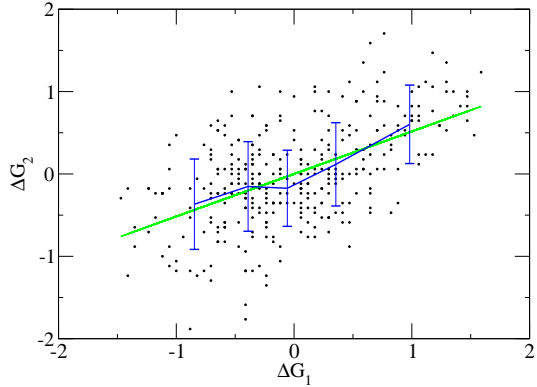}
\caption{\label{deltag_basic} The correlation of $\Delta G$ for
the first and the second half of the season. Included are the
respective averages together with the standard deviation which on
average is 0.51. Furthermore an overall regression line is
included which has a slope of 0.53. }
\end{figure}

A natural question to ask is whether the distribution for $N=34$
can be explained under the assumption that all teams have an
identical fitness. If this is the case the outcome of each match
would be purely statistical and no correlation between the goal
differences of a team in successive matches could be found. To
check this possibility in a simple manner we correlate the value
of $\Delta G$, obtained in the first half of the season ($\Delta
G_1)$, with the value of the second half of the same team ($\Delta
G_2)$. The results, collected for all years and all teams (per
year) are shown in Fig.\ref{deltag_basic}. One observes a
significant correlation. Thus, not surprisingly, there is indeed a
variance of the fitness of different teams.

For a quantification of the correlation one can use the Pearson
correlation coefficient
\begin{equation}
c_P(M_1,M_2) = \frac{<(M_1 - <M_1>)(M_2 - <M_2>)>}{\sigma_{M,1}
\sigma_{M,2}}
\end{equation}
to correlate two distributions $M_1$ and $M_2$.  For the present
problem it yields $0.55 \pm 0.03$. The error bar has been
determined by calculating $c_P(M_1,M_2)$ individually for every
year and then averaging over all years. This procedure is also
applied in most of the subsequent analysis and allows a
straightforward estimation of the statistical uncertainty.  The average
value $\langle \Delta G_2 \rangle$ can be interpreted as the best
estimation of the fitness, based on knowledge of $\Delta G_1$.
Note that the variance of the distribution of $\Delta G_2$ for
every $\Delta G_1$ is basically independent of $\Delta G_1$ and is given by 0.51.

There is a simple but on first view astonishing observation. It
turns out that a team with a positive $\Delta G$ in the first half
will on average also acquire a positive $\Delta G$ in the second
half, but with a smaller average value. This is reflected by the
slope of the regression line smaller than unity. This observation
is a manifestation of the regression toward the mean
\cite{Stigler}, which, however, is not always taken into account
\cite{buch}. Qualitatively, this effect can be rationalized by the
observation that a team with a better-than-average value of
$\Delta G$ very likely has a higher fitness but, at the same time, on
average also had some good luck. This statistical bias is, of
course, not repeated during the second half of the season. For a
stationary process $\Delta G$ has the same statistical properties
in the first and the second half. Then the slope of the regression
line is identical to the correlation coefficient (here: 0.53 vs.
0.55).

In a next step we have taken the observable $p(\Delta G = 2)$
which describes the probability that a team wins a match with a
goal difference of exactly two.  Of course, this is also a measure
of the fitness of the team but intuitively one would expect a
major intrinsic statistical variance which should render this
observable unsuited to reflect the team fitness for the real
situation of a finite season. One obtains a correlation
coefficient of 0.19. In agreement with intuition one indeed
sees that observables which are strongly hampered by
statistical effects display a lower correlation coefficient.
Stated differently, the value of $c_p(M_1,M_2)$ can be taken as a
criterion how well the observable $M$ reflects the fitness of a
team. This statement is further corroborated in Appendix I on the
basis of a simple model calculation. In particular it is shown that this statement holds whether or not the team fitness changes during a season.

We have repeated the analysis for the value of $P$, applying the
present rule (3 points for a win, 1 point for a draw and 0 for a
loss) to all years. The results, however, are basically identical
if using the 2-point rule. Here we obtain $0.49 \pm 0.03$ which is
smaller than the value obtained for $\Delta G$. One might argue
that both values can still agree within statistical errors.
However, since the variation from season to season is very similar
for both correlation factors the difference is indeed significant.
A detailed statistical analysis yields $c_P(\Delta G_1,\Delta G_2)
- c_P(P_1,P_2) = 0.06 \pm 0.015$.

\begin{table}
\centering
  \begin{tabular}[t]{|c|c|}\hline
& $c_p$ \\ \hline
      $\Delta G$  & $0.55\pm 0.035$ \\ \hline
  $P$ & $0.49 \pm 0.035$ \\\hline
 $p(\Delta G)=2$ & $0.19 \pm 0.06$ \\ \hline
       \end{tabular}
       \caption{ Pearson correlation coefficients for different observables.} \label{tab1}
\end{table}

How to rationalize this difference? A team playing 1:0 gets the
same number of points than a team winning 6:0. Whereas in the
first case this may have been a fortunate win, in the second case
it is very likely that the winning team has been very superior. As
a consequence the goal difference may identify very good teams
whereas the fitness variation among teams with a given number of
points is somewhat larger. Actually, using $\Delta G_1$ to predict
$P_2$ is also more efficient than using $P_1$ ($c_P(\Delta
G_1,P_2) > c_P (P_1,P_2)$). One might wonder whether the most
informative quantity is a linear combination of $\Delta G$ and
$P$. Indeed the optimized observable $\Delta G + 0.3 P$ displays a larger
value of $c_P$ than $\Delta G$ alone. The difference, however, is
so small ($\Delta c_P \approx 0.001$) that the additional
information content of the points can be totally neglected.

As a conclusion a final ranking in terms of goals rather than
points is preferable if one really wants to identify the
strongest or weakest teams.

\section{Temporal evolution of the fitness}

Having identified $\Delta G$ as an appropriate measure for the team
fitness one may ask to which degree the team fitness changes with
time. This will be analyzed on three different time scales, now using all data starting
from 1965/66.

\begin{figure}
\includegraphics[width=7cm]{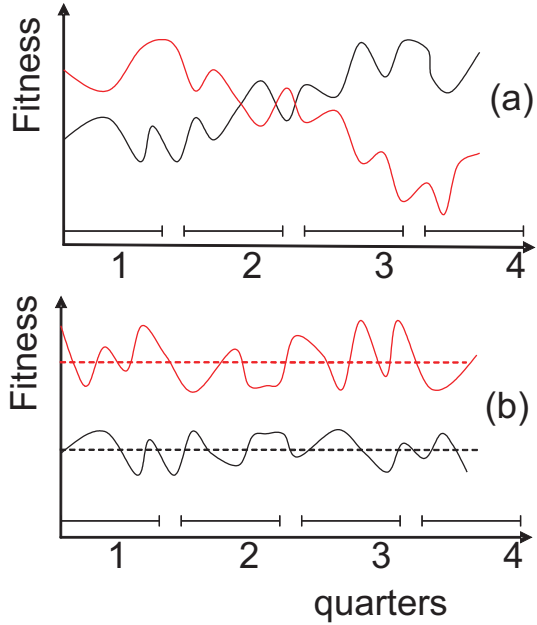}
\caption{\label{sketch_time} Two extreme scenarios for the time
evolution of the fitness during a season. (a) The fitness performs a
random-walk dynamics under the only constraint that the fitness
distribution of all teams is (roughly) stationary. (b) The fitness
of each team fluctuates around a predefined value which is constant
for the whole season. }
\end{figure}

First we start with variations within a season. One may envisage two
extreme scenarios for the time evolution of the fitness during a
season: First a random walk in fitness-space, second fluctuations
around fixed values.  These scenarios are sketched in
Fig.\ref{sketch_time}.

\begin{figure}
\includegraphics[width=7cm]{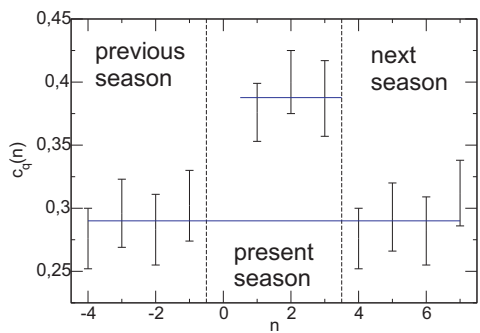}
\caption{\label{viertel_corr} The correlations between quarters,
involving the comparison between subsequent seasons. $n$ denotes the
difference between the quarter indices. For a closer description see
text.}
\end{figure}

To quantify this effect we divide the season in four nearly equal
parts (9 matches, 8 matches, 9 matches, 8 matches), denoted
quarters. The quarters are enumerated by an index from 1 to 4. In
the random-walk picture one would naturally expect that the
correlation of quarters $1$ and $m$ ($m=2,3,4$) is the stronger the
smaller the value of $m$ is. For the subsequent analysis we introduce
the variable $n=m-1$, indicating the time lag between both
quarters. In contrast, in the constant-fitness scenario no
dependence on $n$ is expected. The correlation factors, denoted
$c_q(n)$, are displayed in the central part of
Fig.\ref{viertel_corr}. To decrease the statistical error we have
averaged over the forward direction (first quarter with $m=n+1$-th
quarter) and the time-reversed direction (last quarter with
$m=4-n$-th quarter). Interestingly, no significant dependence on $n$ is observed. The correlation
between the first and the fourth quarter is even slightly larger
than between the first and the second quarter, albeit within the
error bars. Thus, the hypothesis that the fitness remains constant
during a season (apart from short-ranged fluctuations) is fully
consistent with the data. Of course, because of the residual
statistical uncertainties of the correlations, one cannot exclude
a minor systematic variation of the fitness.

This analysis can be extended to learn about a possible fitness
variation when comparing one season with the next or the
previous season. More specifically, we correlate the fitness in
the first quarter of a given season with the quarters $m=5,6,7,8$
in the next season and with the quarters $m=-3,-2,-1,0$ and the
previous season and plot it again as function of $n=m-1$. The
results are also included in Fig.\ref{viertel_corr}.
Interestingly, there is a significant drop of correlation which,
consistent with the previous results, does not change during the
course of the next or the previous season. Thus it is by far the
summer break rather than the time during a season where most
changes happen to the fitness of a team. The very fact that the
correlation to last year's result is weaker than present
year's result has been already discussed in \cite{goddard}, based
on a specific model analysis.

Finally, we have analysed the loss of correlation between seasons
$i$ and $i+n$. In order to include the case $n=0$ in this analysis
we compared $\Delta G$, determined for the first and the second
halves of the season. Thus, for the correlation within the same
season one obtains one data point, for the correlation of
different seasons one obtains four data points which are
subsequently averaged. $c_y(n)$ denotes the corresponding Pearson
correlation coefficient, averaged over all initial years $i$. We
checked that for $n > 0$ we get the same shape of $c_y(n)$ (just
with larger values) when full-year correlations are considered. Of
course, when calculating the correlation coefficient between
seasons $i$ and $i+n$ one only takes into account teams which are
in the Bundesliga in both years. However, even for large time
differences, i.e. large $n$, this number is significant (e.g. the
number of teams playing in the first season, analyzed in this
study, and the season 2007/08 is as large as 11). This already
indicates that, given the large number of soccer teams in Germany
which might potentially play in the Bundesliga, a significant
persistence of the fitness is expected although many of these
teams in between may have been briefly relegated to a lower
league.

\begin{figure}
\includegraphics[width=7cm]{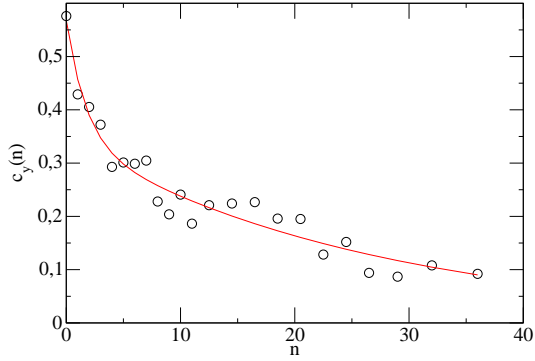}
\caption{\label{jahr_corr2} The fitness correlation when comparing
$\Delta G$ for two seasons which are $n$ years apart. The analysis
is based on the comparison of half-seasons (see text for more
details). The data are fitted by $c_y(n) = 0.22\exp(-n/1.7) + 0.34
\exp(-n/27)$. }
\end{figure}

The results are shown in Fig.\ref{jahr_corr2}. $c_y(n)$ displays a fast decorrelation for
short times which slows down for longer times. To capture these two
time-regimes we have fitted the data by a bi-exponential function (numbers are given in the figure caption).
This choice is motivated by the fact that this is maybe the simplest function
which may quantify the $n$-dependence of $c_y(n)$.
The short-time loss has a time scale of around 2 years. This effect,
however, only has an amplitude of around 2/5 as compared to the
total. The remaining loss of correlation occurs on a much longer
scale (around 20-30 years). Obviously, there exist  fundamental
properties of a team such as the general economic situation which
only change on extremely long time scales given the short-range
fluctuations of a team composition. As mentioned above, this
long-time correlation is also reflected by the small number of
teams which during the last decades have played a significant time
in the Bundesliga.

\section{Statistical description of a soccer league}

\subsection{General}

Here we explicitly make use of the observation that the fitness
does not change during the season. Actually, in this Section we
will report another supporting piece of evidence for this
important fact. Hypothetically, this fitness could be obtained
"experimentally" if a season would contain an infinite number of
matches between the 18 teams. Then, the fitness could be
identified as the observable $\Delta G (N \rightarrow \infty)$
(abbreviated $\Delta G(\infty)$). The specific value for team $i$
is denoted $\Delta G_i (\infty)$.  We already know from the
discussion of Fig.\ref{deltag_basic} that the values $\Delta G_i
(\infty)$ are distributed. As a consequence the variance of
$\Delta G(\infty)$, denoted $\sigma^2_{\Delta G}$, is non-zero.
Although it cannot be directly obtained from the soccer table
(because of the finite length of a season) it can be estimated via
appropriate statistical means, as discussed below.  Because the
number of goals and the width of the distribution of $\Delta G$
somewhat decreased if comparing the years starting from the season
1987/88 with the earlier years, we restrict the analysis in this
section to the latter time regime.

\subsection{Estimation of the statistical contribution}

Formally, the omnipresence of statistical effects can be
written as
\begin{equation}
\label{gn_def} \Delta G_i(N) = \Delta G_i(\infty) + \Delta
G_{i,stat}(N).
\end{equation}
In physical terms this corresponds to the case of a biased random
walk, i.e. a set of particles, each with a distinct velocity
(corresponding to $(\Delta G_i(\infty))$) and some diffusion
contribution (corresponding to $\Delta G_{i,stat}(N)$). We note in
passing that to a good approximation the amplitude of the
statistical contribution does not depend on the value of the
fitness, i.e. the index $i$ in the last term of Eq.\ref{gn_def}
can be omitted. Otherwise, the variance in Fig.\ref{deltag_basic}
would depend on the value of $\Delta G_1$. 

Squaring Eq.\ref{gn_def} and averaging over all teams one can write
\begin{equation}
\label{statsum} \sigma^2_{\Delta G (N)}  = \sigma^2_{\Delta
G}  + \sigma^2_{\Delta G (N),stat}
\end{equation}
where the variances of the respective terms have been introduced.
 $\sigma^2_{\Delta
G (N),stat}$ is expected to scale like $1/N$ and will disappear in
the limit $N \rightarrow \infty$. Thus, $\sigma^2_{\Delta G}$ can
be extracted by linear extrapolation of $\sigma^2_{\Delta G (N)}$
in a $1/N$-representation. We have restricted ourselves to even
values of $N$ in order to avoid fluctuations for small $N$ due to
the differences between home and away matches. To improve the
statistics we have not only used the first $N$ matches of a season
but used all sets of $N$ successive matches of a team for the
averaging. This just reflects the fact that any $N$ successive matches have the same information content about
the quality of a team.

\begin{figure}
\includegraphics[width=7cm]{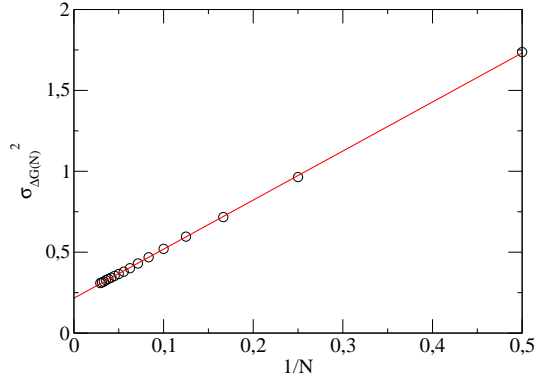}
\caption{\label{fitness_nall} The variance of the distribution of
$\Delta G(N)$, averaged over all years. The straight line is a
linear fit.  }
\end{figure}

One can
clearly see in
Fig.\ref{fitness_nall} that one obtains a straight line in the $1/N$-representation for all values of $N$.
We obtain
\begin{equation}
\label{eqs2} \sigma^2_{\Delta G(N)} = 0.215 +
\frac{3.03}{N},
\end{equation}
i.e.  $\sigma^2_{\Delta G} = 0.215$  and $\sigma^2_{\Delta G
(N),stat}= 3.03/N$.  Generally speaking, the excellent linear fit
in the $1/N$-representation shows again that the team fitness
remains stable during the season.
Otherwise one would expect a
bending because also the first term in Eq.\ref{statsum} would
depend on $N$; see again Appendix I for a more quantitative
discussion of this effect. Of course, for this statement it was important to include
only {\it successive} matches of a team for the statistical analysis.

\begin{figure}
\includegraphics[width=7cm]{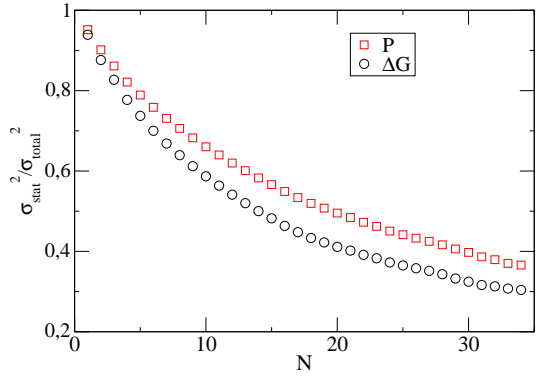}
\caption{\label{fitness_stat} Statistical contribution to the
overall variance after $N$ matches. Included is the analysis for the
goal differences as well as for the points.}
\end{figure}

In Fig.\ref{fitness_stat} the relative contribution of the
statistical effects in terms of the variance, i.e.
$\sigma^2_{\Delta G (N),stat}/( \sigma^2_{\Delta G (N),stat} +
\sigma^2_{\Delta G })$ is shown as a function of $N$. The result
implies that, e.g., after the first match of the season ($N=1$)
approx. 95\% of the overall variance is determined by the
statistical effect. Not surprisingly, the table after one match
may be stimulating for the leading team but has basically no
relevance for the rest of the season. For $N \approx 14$ the
systematic and the statistical effects are the same.
Interestingly, even at the end of the season the statistical
contribution in terms of its contribution to the total variance is still as large as 30\%.

Repeating the same analysis for the number of points $P$ yields
\begin{equation}
\label{point}
 \sigma^2_{P(N)}\approx 0.08 +
\frac{1.7}{N}.
\end{equation}
The resulting plot of $\sigma^2_{P(N),stat}/( \sigma^2_{P(N),stat} +
\sigma^2_{P})$ is again displayed in Fig.\ref{fitness_stat}. Now it takes even $N=22$ matches
until the systematic effects start to be dominant. At the end of the season the
statistical contribution is as large as 36\%. This shows
again that $\Delta G$ is a better measure for the fitness
because then the random component in the final ranking is somewhat
smaller.

\subsection{Prediction of team fitness: General framework}

The previous analysis has shown that even for $N=34$ there still
exists a significant random contribution. The next goal is to
estimate in a statistically consistent way from knowledge of $\Delta
G(N)$ (e.g. the final scores at the end of the season) the team
fitness. Formally, one wants to determine the conditional
probability function $p(\Delta G(\infty) | \Delta G(N))$. This can
be determined by using the Bayes theorem
\begin{equation}
\label{Bayes} p(\Delta G(\infty) | \Delta G(N)) \propto p(\Delta
G(N) | \Delta G(\infty))) q(\Delta G(\infty))
\end{equation}
Here $p(\Delta G(N) | \Delta G(\infty))$ is fully determined via
Eq.\ref{gn_def} and corresponds to a Gaussian with variance
$\sigma^2_{\Delta G (N),stat}$. The function $q(\Delta G(\infty))$
describes the a priori probability for the team fitness. This
distribution has been already discussed in  Fig.\ref{deltag_dist}.
To first approximation we saw a Gaussian behavior with small but
significant deviations. One can show that a strict linear
correlation between the estimated fitness (or the behavior in the
second half of the season) and $\Delta G(N)$ is fulfilled for a
Gaussian distribution $q(\Delta G(\infty))$. Since to a good
approximation a linear correlation was indeed observed in
Fig.\ref{deltag_basic}, for the subsequent analysis we neglect any
deviations from a Gaussian by choosing $q(\Delta G(\infty))
\propto \exp(-\Delta G(\infty)^2/2\sigma^2_{\Delta G})$. Of
course, for a more refined analysis the non-Gaussian nature,
displayed in Fig.\ref{deltag_dist}, could be taken into
account.

After reordering of the Gaussians in Eq.\ref{Bayes} one obtains
after a  straightforward calculation
\begin{equation}
\label{cond}  p(\Delta G(\infty) | \Delta G(N)) \propto
\exp[-(\Delta G(\infty) - a_N \Delta G(N))^2/2\sigma^2_{e,N}).
\end{equation}
with
\begin{equation}
\label{pred_a} a_N = \frac{\sigma^2_{\Delta G}
}{\sigma^2_{\Delta G} + \sigma^2_{\Delta G(N),stat}}
\end{equation}
and
\begin{equation}
\sigma_{e,N}^2 = \frac{\sigma^2_{\Delta G (N),stat}}{1+
\sigma^2_{\Delta G (N),stat}/\sigma^2_{\Delta G }}.
\end{equation}
As discussed in the context of Fig.\ref{deltag_basic} $a_N$ is
identical to the Pearson correlation coefficient when correlating
two subsequent values of $\Delta G$, each based on $N$ matches.

From Eq.\ref{eqs2} one obtains $a_{N=17} = 0.55$ and
$\sigma_{e,N=17}^2 = 0.097$. As expected $a_N$ is identical to
$c_P(\Delta G_1,\Delta G_2)$ and within statistical uncertainties
identical to the slope of 0.53 in Fig.\ref{deltag_basic}.

Finally, we apply these results to the interpretation of the
Bundesliga table at the end of the season, i.e. for $N=34$. Using
Eq.\ref{cond} the estimator for $\Delta G(\infty)$  can be written
as
\begin{equation}
\Delta G(\infty) = a_{N=34} \Delta G(N=34) \pm \sigma_{e,N=34}.
\end{equation}
 For the
present data this can be explicitly written as
\begin{equation}
\label{Gest} \Delta G(\infty) = 0.71 [\Delta G(N=34) \pm 0.36] .
\end{equation}
Using standard statistical analysis one can, e.g., determine the
probability that a team with a better goal difference $\Delta G$
(i.e. $\Delta G_1> \Delta G_2)$ is indeed the better team. For the
present data it turns out that for $\Delta G_1 - \Delta G_2 =
0.36$ (corresponding to an absolute value of 12 goals after 34
matches) the probability is approx. 24\% that the team with the
worse goal difference is nevertheless the better team.

In analogy, one can estimate from Eq.\ref{point} that two teams which after the season
are 10 points apart have an incorrect order in the league table, based on their
true fitness, with a probability of 24\%. Maybe this figure more dramatically reflects
the strong random component in soccer.

\subsection{Prediction of team fitness: Application}

These results can be taken to quantify the uncertainty when
predicting $\Delta G_i(M)$ of team $i$. More specifically, we
assume that this prediction is based on the knowledge of the
results of the $N$ previous
 matches of team $i$. The variance of the estimate of $\Delta G(M)$ is denoted
 $\sigma^2_{est}(M,N)$. This notation reflects the fact that it depends
 on both the prediction time scale $M$ as well as the information time scale $N$.
To estimate $\Delta G_i(M)$, based on $\Delta G_i(N)$, two
uncertainties have to be taken into account. First, the
uncertainty of estimating $\Delta G_i(\infty)$ is characterized by
$\sigma^2_{e,N}$. Second, even if $\Delta G( \infty)$ were
known exactly, the statistical uncertainty of estimating $\Delta G(M)$ due to the finite $M$
is still governed by the variance $\sigma^2_{\Delta G
(M),stat}$. Thus, one obtains
\begin{equation}
\label{pred} \sigma^2_{est}(M,N) = \sigma^2_{e,N} +
\sigma^2_{\Delta G (M),stat}
\end{equation}

\begin{figure}
\includegraphics[width=7cm]{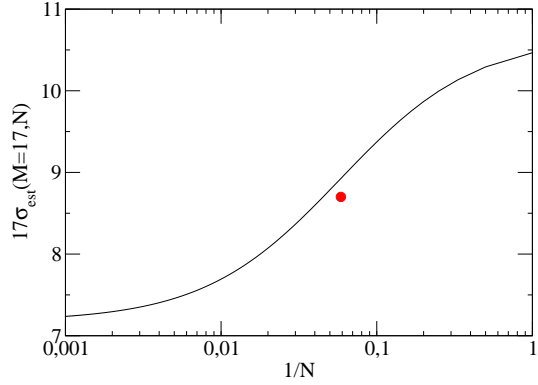}
\caption{\label{prediction} The function $17
\sigma_{est}(M=17,N)$, describing the uncertainty for the
prediction of the goal difference during the second half of the
season based on the knowledge of $N$ matches. Included is the data
point, observed numerically in Fig.\ref{deltag_basic}.}
\end{figure}

For the specific choice $M=17$, i.e. for the prediction of the
second half of the season, the standard deviation $17 \cdot
\sigma_{est}(M=17,N)$ of the estimator (expressed in absolute
number of goals) is displayed in Fig.\ref{prediction}.
First, we discuss the extreme cases. In the practically impossible
case that the fitness is exactly known (formally corresponding to
$N \rightarrow \infty$) one obtains a standard deviation of approx. 7 goals.
In the other extreme limit where no information is available, i.e.
$N = 0$) one obtains a value of approx. 10.5 goals. Thus
the difference between complete information and no information for
the prediction of the second half of the season is only 3.5 goals.
Finally, for the interpretation of the results in
Fig.\ref{deltag_basic} one has to choose $N=17$. As shown in
Fig.\ref{prediction} the observed standard deviation of $17\cdot
0.51 \approx 8.7$ agrees  well with the theoretical value based on
Eq.\ref{pred}. The remaining deviations (8.7 vs. 8.9) might reflect the
non-Gaussian contributions to $q(\Delta G(\infty))$.

From Eq.\ref{point} one can estimate in analogy to above
that, based on the knowledge of the points for the first half, the number
of points for the second half can be estimated with a standard deviation
of approx. 6 points. Of course, according to our previous discussion the
estimation would be slightly better if the value of $\Delta G$ rather
than the number of points of the first half were taken as input.

\subsection{Going beyond the team fitness $\Delta G$}

So far we have characterized the fitness of a team $\Delta G$.
From a conceptual point of view the most elementary quantities are
the number  of goals $G_+$, scored by a team, as well as the
number of goals $G_-$ conceded by this team ( $\Delta G = G_+ -
G_-$). Correspondingly, $\langle G_\pm \rangle $ denotes the
average number of goals per team and match. The brackets denote
the corresponding average. Since the subsequent analysis can be
also used for prediction purposes we restrict ourselves to all
years since the season 1995/96 when the 3-point rule had been
introduced.

The above analysis, performed for $\Delta G$, can be repeated
for $G_\pm$. The general notation reads ($M \in \{G_+,G_-$\})
\begin{equation}
\sigma^2_{M(N)} = \sigma^2_{M} + \frac{b_M}{N}.
\end{equation}

The fitting parameters are listed in Tab.II.  We note in passing
that all statistical features, described in this Section, are
observed in the English Premier League, too. For reasons of
comparison the resulting parameters are also included in Tab.II.
\begin{table}
\label{tab2}
\centering
  \begin{tabular}[t]{|l|c|c|c|c|c|c|}\hline &
$\langle G_\pm \rangle$  & $\sigma^2_{G_+}$& $b_{G_+}$ & $\sigma^2_{G_-}$&
$b_{G_-}$ & $c_{+,-}$\\
\hline
      Bundesliga & 1.43   & 0.075 & 1.45 & 0.055 & 1.50 & 0.71 \\ \hline
     Premier League & 1.29  &  0.075 & 1.40 & 0.060 & 1.40 & 0.85 \\ \hline
       \end{tabular}
       \caption{ Statistical parameters, characterizing the Bundesliga (1995/96-2007/08) and the English
       Premier League (1996/97-2006/07).
       }
\end{table}

For a complete understanding of the goal statistics one has to include possible correlations between $G_+$ and
$G_-$, i.e.
\begin{equation}
c_{+,-}(N) = \frac{\langle (G_+ - \langle G \rangle) ( \langle G
\rangle - G_-)}{\sigma_{G_+}\sigma_{G_-}} .
\end{equation}
This value reflects the correlation of a team's strength of attack and  defence. Complete correlation means $c_{+,-}(N) = 1$.
The statistical effects during a soccer match, related to $G_+$ and $G_-$, are likely to be statistically uncorrelated.
As a consequence one would not expect a significant $N$-dependence. Indeed, we have verified this expectation by explicit calculation
of $c_{+,-}(N)$ which within statistical uncertainty is $N$-independent. We obtain $c_{+,-} = 0.71$.

This information is sufficient to calculate $\sigma^2_{M(N)}$ for $M\in \{\Delta G \equiv G_+ - G_-,\Sigma G \equiv G_+ + G_-\}$ via
 $\sigma^2_{(G_+ \pm G_-)(N)} = \sigma_{G_+(N)}^2 +  \sigma_{G_-(N)}^2 \mp 2c_{+,-} \sigma_{G_+}\sigma_{G_-}$. One obtains
 $\sigma^2_{\Delta G}(N) = 0.22 + 2.95/N$ and  $\sigma^2_{\Sigma G}(N) = 0.03 + 2.95/N$.  $\sigma^2_{\Delta G}(N)$ agrees very well
 with the data, reported above for the time interval 1987/88-2007/08.

Based on this detailed insight into the statistical nature of
goals several basic questions about the nature of soccer can be
answered.

Are offence or defence abilities more important? The magnitude of
the variance $\sigma^2_{M}$ is a direct measure for the relevance
of the observable $M$. Since $\sigma^2_{G_+} / \sigma^2_{G_-}
=1.25 \pm 0.09 > 1$ the investment in good strikers may be
slightly more rewarding. However, the difference is quite small
 so that to first approximation both aspects of a soccer match are of similar importance.

Do teams with good strikers also have a good defence? In case of a strict correlation one would have $c_{+,-} = 1$. The present value of 0.71 indicates that there
  is indeed a strong correlation. However, the residual deviation from unity reflects some team dependent differences beyond simple statistical fluctuations.
  Interestingly, this correlation is significantly
  stronger in the Premier League, indicating an even stronger balance between the offence and the defence in a team of the Premier League.

Is the total number of goals of a team (i.e. $G_+ + G_-$) a
team-specific property? On average this sum is 97. Without
statistical effects due to the finite length of a season the standard
deviation of this value would be just $ 34 \sigma_{\Sigma G }
\approx 6$, i.e. only a few percent. Thus, to a very good approximation the number of goals on average scored by team $i$
is just given by $G_{+,i} = \langle G_\pm \rangle + \Delta G_i/2$ (an analogous formula holds for $G_{-,i}$).

\section{Soccer myths}

In typical soccer reports one can read that a team is particularly
strong at home (or away) or is just playing a winning streak ({\it
Lauf} in German) or a losing streak. Here we show that the actual
data does not support the use of these terminologies (except for
the presence of losing streaks).

\subsection{Home fitness}

One may ask the general question whether the overall fitness $\Delta G$ of the team {\it fully} determines the {\it home
fitness}, i.e. the team quality of playing at home. If yes, it would be useless and misleading to define
a team-specific home fitness because it is not an independent observable but
just follows from the overall fitness $\Delta G(\infty)$. For the present analysis we use again our standard data set
starting from 1987/88.

To discuss the ability of a team to play at home as compared to play
away we introduce $\Delta G_H(N)$ and $\Delta G_A(N)$ as the goal
difference in $N$ home matches and $N$ away matches, respectively. Of course,
one has $\Delta G_H (N) + \Delta G_A (N) = \Delta G (2N)$. The {\it home
advantage} can be characterized by
\begin{equation}
\Delta (\Delta G) = \Delta G_H - \Delta G_A.
\end{equation}
The average value $\langle \Delta (\Delta G) \rangle $ is approx. 1.4,
which denotes the improved home goal difference as compared to the
away goal difference. This number also means that on average a team
scores 0.7 more goals at home rather than away whereas 0.7 goals
more are conceded by this team when playing away.  We note in passing
that the home advantage is continuously decreasing with time. Just taking
the seasons since 1995/96 one gets, e.g., $\Delta (\Delta G) \approx 1.0$.

A team-specific home fitness could be characterized by $\Delta
(\Delta G)_i - \langle \Delta (\Delta G)\rangle $. A positive
value means that team $i$ is better at home than expected from the
overall fitness $\Delta G$. Of course, again one has to consider
the limit $N \rightarrow \infty$. Thus, in analogy to the previous
Section one has to perform a scaling analysis. After $N$ matches
$\Delta (\Delta G)(N)$ will be distributed with a variance,
denoted $\sigma^2_{\Delta (\Delta G) (N)}$.  A positive value of
the large $N$-limit $\sigma^2_{\Delta (\Delta G)}$ reflects the
presence of a home fitness.  Otherwise the quality of a team for a
match at home (or away) is fully governed by the overall fitness
$\Delta G(\infty)$.

\begin{figure}
\includegraphics[width=7cm]{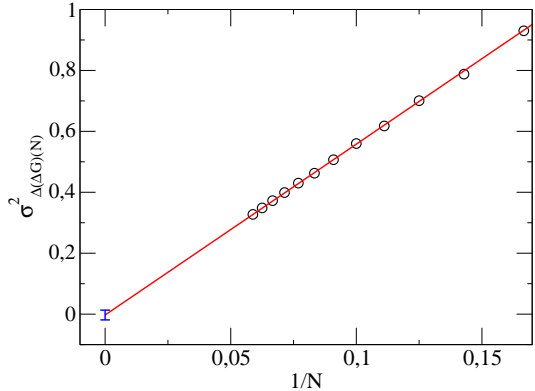}
\caption{\label{heimst_n} The variance of $\Delta (\Delta G)$, i.e.
$\sigma^2_{\Delta (\Delta G(N))}$ vs. $1/N$. The straight line is a
linear fit. The extrapolation to $N=\infty$ yields
approx. $-0.003 \pm 0.016$.}
\end{figure}

The $N$-dependence of $\sigma^2_{\Delta (\Delta G) (N)}$ is shown
in Fig.\ref{heimst_n}. To obtain these data one has to evaluate
the appropriate expression for the empirical variance for this
type of analysis which is a slightly tedious but straightforward
statistical problem. The statistical error has been estimated from performing
this analysis for the individual years.

It becomes clear that the hypothesis $\sigma^2_{\Delta (\Delta G)
(\infty)} =0$ is fully compatible with the data. Because of the
intrinsic statistical error one cannot exclude a finite value of
$\sigma_{\Delta (\Delta G) (\infty)}$ ($\sigma_{\Delta (\Delta G)
(\infty)}  < 0.12)$.  This value is less than 10\% of the
average value $\langle \Delta (\Delta G) \rangle = 1.4$. Thus, the presence of teams which are
specifically strong at home relative to their overall fitness is, if at all,
a very minor effect.

Although this result rules out the presence of a relevant team-specific home fitness
it may be illuminating to approach the same problem from a direct analysis of
the whole distribution of $\Delta (\Delta G)(N=17)$. The goal is to compare it with the distribution
one would expect for the ideal case where no team-specific home fitness is present. This
comparison, which is technically a little bit involved, is shifted
to Appendix II. It turns out that the residual home fitness can be
described by a value of $ 0 \le \sigma_{\Delta
(\Delta G)}\ll 0.4$. This means
that in particular the simple model, sketched above, is not
compatible with the data. In summary, relative to the average home advantage
of 1.4 any possible residual home fitness is a negligible effect.

In literature it is often assumed that for a specific match of
team A vs. team B one can a priori define the expectation value of
goals $t_{A(h)}$ and $t_{B(a)}$, scored by the home team A and the away team
B, respectively. In the approach of Ref.\cite{Rue00} one
explicitly assumes $t_A(h) = f_{AB} \cdot c_h$ and $t_B(a) =
f_{BA} \cdot c_a$ (using a different notation). Here
$f_{ij}$ contains the information about the offence strength of
team $i$ and the defence strength of team $j$. The information
about the location of the match is only incorporated into the factors
$c_h$ and $c_a$. This approach has two implicit assumptions.
First, the fact that $c_h$ is team-independent is equivalent to
the assumption that there is no team-specific home fitness. This
is exactly what has been shown in this Section. Second, the
average number of goals of, e.g., the home team is proportional to
the average number, expected in a neutral stadium. For reasons of
convenience this number can be chosen identical to $f_{AB}$. Then,
$c_h > 1$ takes into account the general home advantage. The same
holds for $c_a < 1$. Assuming the multiplicative approach one has
to choose
\begin{equation}
c_{h,a} = \frac{\langle G_{\pm} \rangle \pm \langle \Delta (\Delta G)\rangle }{\langle G_{\pm} \rangle}.
\end{equation}
which for the present case yields $c_h/c_a \approx 1.45$

In principle, one might have also added some fixed value to take
into account the home advantage. Thus, the multiplicative approach
is not unique.  However, using the above concepts, one can show
that this approach is indeed compatible with the data. For this
purpose we introduce the observables $M\in \{G_{+,h},G_{+,a},
G_{-,h}, G_{-,a}\}$. $G_{\pm,h}$ denotes the number of goals
scored and conceded by the home team. An analogous definition
holds for $G_{\pm,a}$. In analogy to above one can calculate
$\sigma^2_{M}$ obtained again from the $N \rightarrow
\infty$-extrapolation of the respective observable. One obtains
$\sigma^2_{G_{+,h}} = 0.089 , \sigma^2_{G_{+,a}} = 0.044
,\sigma^2_{G_{-,h}} = 0.033, \sigma^2_{G_{-,a}} = 0.069$. If the
properties of home and away goals are fully characterized by the
factors $c_{h,a}$ one would expect
$\sigma_{G_{+,h}}/\sigma_{G_{+,a}} =
\sigma_{G_{-,a}}/\sigma_{G_{-,h}} = c_h/c_a$. The two ratios read
1.4 and 1.45, respectively, and are thus fully compatible with the
theoretically expected value of 1.45. In case of an additive
constant to account for the home advantage one would have expected
a ratio of 1 because then the distributions would have been just
shifted to account for the home advantage.

In practical terms this allows one to correct the results of
soccer matches for the home advantage by dividing the number of
goals in a match by $c_h$ and $c_a$, respectively. This correction
procedure may be of  interest in cases where one wants to identify
statistical properties without being hampered by the residual home
advantage. Using this procedure for the data points in Fig.6 the
data points for odd $N$ would also fall on the regression line. We
just mention in passing that in the limit of small $\langle
\Delta(\Delta G)\rangle/\langle G_\pm \rangle$ and small
$\sigma_{G_\pm} /  \langle G_\pm \rangle$ (which in practice is
well fulfilled) this scaling yields similar results as compared to
a simple downward shifting of the home goals and upward shifting
of the away goals by $\langle \Delta(\Delta G)\rangle /2$.

\subsection{Streaks}

The aspect of identifying winning or losing streaks is somewhat
subtle because one has to take care that no trivial selection
effects enter this analysis. Here is one example of such an
effect. Evidently, in case of a winning streak it is likely that
during this period the team played against somewhat weaker teams
and will, subsequently, on average play against somewhat stronger
teams. Thus, to judge the future behavior of this team one needs a
method which takes these effects in a most simple way into
account. To obtain a sufficiently good statistics here we use our
complete data set, starting from the season 1965/66.

\begin{figure}
\includegraphics[width=7cm]{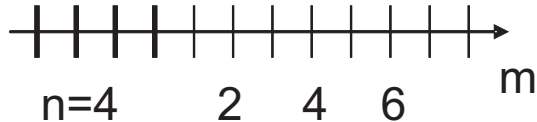}
\caption{\label{sketch_series} Sketch of the definitions of $n$
and $m$ for the analysis of the possible existence of winning and
losing streaks. }
\end{figure}

The key question to be answered here is whether or not the
presence of a winning or losing sequence stabilizes or
destabilizes a team or maybe has no effect at all. If a winning
sequence stabilizes a team one may speak of a winning streak.
Analogously, if a losing sequence destabilizes a team one has a
losing streak. In general, we have identified all sequences of $n$
successive matches where $n$ wins or losses were present. Of
course, the actual length of the win or loss sequences can have
been much longer. Having identified such a sequence we have
determined the probability that in the $m$-th match after this
sequence that team will win. This probability is denoted
$p_{win}(m,n)$. This is sketched in Fig.\ref{sketch_series} for
the case $n=4$.

\begin{figure}
\includegraphics[width=7cm]{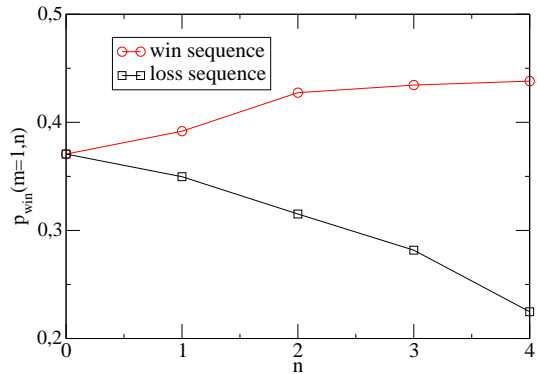}
\caption{\label{series_simple} The probability $p_{win}(n,m)$. to
win after a team as won or lost n times.}
\end{figure}

In a first step we analyze the winning probability in the next
match, i.e. for $m=1$.  The data are shown in
Fig.\ref{series_simple}. In case of winning sequence the
probability to win increases with increasing $n$. The opposite
holds for a losing sequence. Does this indicate that the longer
the winning (losing) sequence, the stronger the (de)stabilization
effect, i.e. real winning or losing streaks emerge?

This question has been already discussed in Ref. \cite{Dobson03}. It
was correctly argued that by choosing teams which have, e.g., won 4
times one typically selects a team with a high fitness. This team
will, of course, win with a higher probability than an average team
(selected for $n=0$). Thus the increase of the win probability with
$n$ is expected even if no stabilizing effect is present. It would
be just a consequence of the presence of the fitness distribution
and thus of good and bad teams, as shown above. Only if all teams
had the same fitness the data of Fig.\ref{series_simple} would
directly indicate the presence of a stabilization and
destabilization effect, respectively.

The key problem in this analysis is that the different data points
in Fig.\ref{series_simple} belong to different subensembles of teams
and thus cannot be compared. Therefore one needs to devise an
analysis tool, where a fixed subensemble is taken. The realization
of this tool is inspired by 4D NMR experiments, performed in the 90s
in different groups to unravel the properties of supercooled liquids
\cite{Klaus,Wilhelm,epl}. The key problem was to monitor the time
evolution of the properties of a specific subensemble until it
behaves again like the average. This problem is analogous to that of
a soccer team being selected because of $n$ wins or losses in a row.

This idea can be directly applied to the present problem by
analyzing the $m$-dependence of $p_{win}(m,n)$.  It  directly
reflects possible stabilization or destabilization effects. In
case of a stabilization effect $p_{win}(m)$ would be largest for
$m=1$ and then decay to some limiting value which would be related
to the typical fitness of that team after possible effects of the
series have disappeared. In contrast, in case of a destabilization
effect $p_{win}(m=1)$ would be smaller than the limiting value
reached for large $m$. Note that in this way the problem of
different subensembles is avoided. Furthermore this analysis is
not hampered by the fact that most likely the opponents during the
selection period of $n$ matches were on average somewhat weaker
teams. The limiting value has been determined independently by
averaging $p_{win}(m,n)$ for $|m| > 8$, i.e. over matches far away
from the original sequence. To improve the statistical quality
this average also includes the matches sufficiently far before the
selected sequence (formally corresponding to negative $m$). Of
course, only matches within the same season were taken into
account. It is supposed to reflect the general fitness of a team
during this season (now in terms of wins) independent of that
sequence. In case of no stabilization or destabilization effect
the observable $p_{win}(m,n)$ would not depend on $m$. This would
be the result if playing soccer would be just coin tossing without
memory. To avoid any bias with respect to home or away matches we
only considered those sequences where half of the matches were
home matches and and the other half away matches ($n$ even).
Furthermore, the data for $p_{win}(m,n)$ are averaged pairwise for
subsequent $m$ (1 and 2, 3 and 4, and so on).

\begin{figure}
\includegraphics[width=7cm]{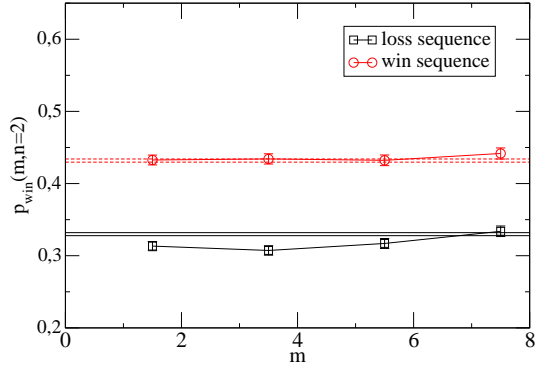}
\caption{\label{series2} The probability to win $p_{win}(m,n=2)$
after a sequence of $n=2$ wins and losses, respectively. The broken
lines indicate the range ($\pm 1\sigma$-interval) of the plateau
value reached for large $m$.}
\end{figure}

\begin{figure}
\includegraphics[width=7cm]{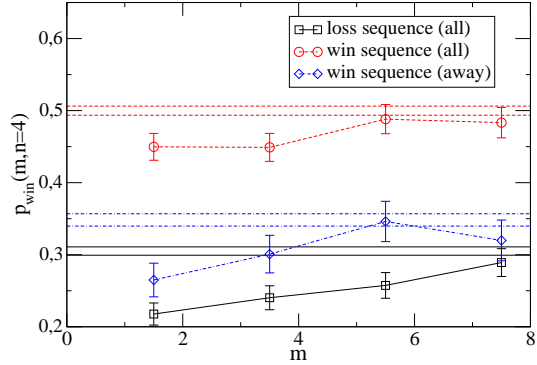}
\caption{\label{series4} Same as in the previous figure for $n=4$.
In addition we have included data where only away matches of the
teams are considered for the calculation of $p_{win}(m,n=4)$ in case
of a win sequence. }
\end{figure}

The functions $p_{win}(m,n)$ for $n=2$ and $n=4$ are shown in Figs.
\ref{series2} and \ref{series4}, respectively. For $n=4$ a total of 374 win
sequences and 384 loss sequences have been taken into account. For
$n=2$ one observes a small but significant destabilization after a
loss sequence. It takes approx. 8 matches to recover. No effects are
seen for the win sequence. More significant effects are visible for
$n=4$. For the loss sequence one observes that directly after the
selected sequences, i.e. for $m=1$ and $m=2$ the winning probability
is reduced by approx. 30\% as compared to the limiting value. Thus
for about 6 matches the teams play worse than normal. Surprisingly,
a reduction of $p_{win}(m,n=4)$ for small $m$ is also visible for
the win sequence. Thus, there seems to be a destabilization rather
than a stabilization effect. By restricting the analysis to the away
matches after the selected sequence, this effect is even more
pronounced. Of course, correspondingly the effect is smaller for
home matches. Unfortunately, $n=6$ can no longer be analyzed because
due to the small number of events the statistics is too bad.

\begin{figure}
\includegraphics[width=7cm]{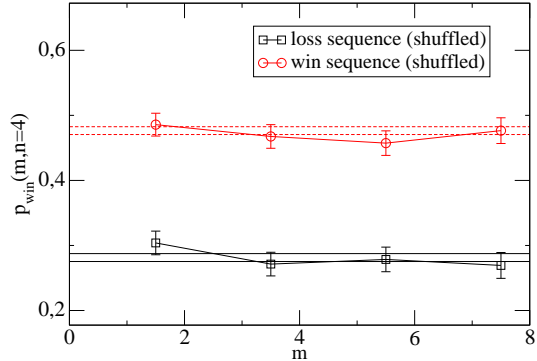}
\caption{\label{series4_shuffle} Analysis of loss and win sequences, using
shuffled data.}
\end{figure}

Of course, a critical aspect in this discussion is the matter of
statistical significance. For this purpose we have estimated the
probability that, using Gaussian statistics, the average of the
first four matches after a win sequence can be understood as an
extreme statistical deviation from the final plateau value. This
probability turns out to be smaller than $10^{-3}$. Furthermore we
analyzed shuffled data, i.e. where for a given team in a given
season the 34 matches are randomly ordered. The results for
$p_{win}(m,n=4)$, using one example of ordering, are shown in
Fig.\ref{series4_shuffle}. As expected no effect is seen. The
observation that the plateau values are somewhat lower than in
Fig.\ref{series4} just reflects the fact the the first data points
(small $m$) in Fig.\ref{series4} are systematically lower than the
respective plateau value.

Thus, we conclude that both a positive ($n = 4$) as well as a
negative sequence $(n = 2,4)$ have a destabilizing effect. This
means that losing streaks indeed exist whereas there are no
stabilization effects for positive sequences, invalidating the
notion of a winning streak. Rather destabilization effects occur
after a longer winning sequence. This asymmetry between positive
and negative sequences is already reflected by the asymmetry, seen
in Fig.\ref{series_simple}.

Actually, the present results disagree with the statistical
analysis in Ref.\cite{goddard01} for the Premier League. In that
work it is concluded that sequences of consecutive results tend to
end sooner than they should without statistical association.
However, the presence of losing streaks has been clearly
demonstrated above. The disagreement might be due to the different
data set (Bundesliga vs. Premier League). However, one needs to
take into account that in that work the results have been obtained
within a framework of a specific model via Monte-Carlo
simulations. The present analysis has the advantage that, first,
it does not to refer to any model about the nature of soccer and,
second, can be done without additional Monte-Carlo simulations.
Thus, possible artifacts of the model might hamper the
interpretation of the data.

\section{Discussion and Summary}

On a conceptual level we have used finite-size scaling methods to extract the
underlying distribution of fitness parameters. It turns out that the
goal difference is a better measure of the team fitness than the
 number of points. From a technical point of view a key aspect was to analyze
the $N$-dependence of observables such as $\Delta G$. This
problem is analogous to the simple physical problem of  random
walks with a drift. The key results can be summarized as follows.

1.) The fitness of a team displays a complex temporal evolution.
Within a season there are no indications for any variations (except
maybe for day-to-day fluctuations around some average team fitness
which can only be identified via a single-match analysis. This
is, however, beyond the scope of the present work). During the
summer-break a significant decorrelation is observed. This
short-scale decorrelation stops after around 2 years where approx.
40\% of the fitness has been changed (some teams becoming better,
some worse). Interestingly, the remaining 60\% of the fitness only
decorrelates on an extremely long times scale of 20-30 years which is
 close to the data window of our analysis. This shows that
there are dramatic persistence effects, i.e. there are some
underlying reasons why good teams remain good on time scales largely
exceeding the lifetime of typical structures in a club (manager, coach, players
etc.).

2.) For finite seasons (which, naturally, is realized in the actual
soccer leagues) the fitness of a team can be only roughly estimated
because of the presence of residual statistical fluctuations.
However, by linear extrapolation of the variance of the team fitness
one can identify the underlying
variance one would (hypothetically) obtain for an infinite number of
matches. Based on this one can estimate the statistical contribution
to the end-of-the-season table which is quite significant (36\% for points).
This allows one to quantify, e.g., the relevance of the final league
table in some detail. 

3.) The overall fitness, defined via the goal difference $\Delta G$,
is to a large extent the only characteristics of a team. In
particular there is no signature of the presence of a team specific home
fitness. We would like to stress that the definition of a home
fitness is always relative to a single season. This means if a team
is strong at home in one year and weak in another year this would
nevertheless show up in the present analysis. Whenever a team plays
better or worse at home than expected (measured via $\Delta G_H$ - $\Delta
G_A$) this effect can be fully explained in terms of the natural
statistical fluctuations, inherent in soccer matches.

4.) A more detailed view on the number of goals reveals that the
quality of the offence and that of the defence of a team is
strongly correlated. In case of a perfect correlation their
quality would be fully determined by the overall fitness. However,
since the correlation is not perfect there exist indeed
differences. Furthermore,
the strength of attack is slightly more important for a successful
soccer team than the strength of defence, albeit the difference is
not big.

5.) It is possible to identify the impact of the home-advantage
for the final result.  Stated differently, one can estimate the
average outcome of a match one would obtain at a neutral stadium.
This procedure may be helpful if data are taken as input for a
statistical analysis.

6.) The notion of streaks, as present in the soccer language, can
only be confirmed in case of a losing streak. This means that if a
team has lost several times (we analyzed 2 and 4 times) there is a
significant drop of their fitness as compared to the normal level
which will be reached again sufficiently far away from the time
period. Possible reasons may be related to psychological aspects
as well as the presence of persistent structural problems (such as
heavily injured players). Surprisingly, no winning streak could be
identified. Winning two times had no effect on the future outcome.
Winning four times even reduced the fitness, in particular when
having an away match. This analysis had to be performed with care
in order to avoid any trivial statistical effects.  Possibly, this
indicates an interesting psychological effect. In literature one
can find models for understanding the basis of human motivation.
In one of the standard models by Atkinson a reduction of
motivation may occur if the next problem {\it appears} either to
be too difficult (after having lost several times) or too simple
(after having won several times) \cite{atkinson}. However, since
these types of sequences (for $n=4)$  of wins or losses are
relatively rare they are of very minor relevance for the overall
statistical description of the temporal evolution of soccer
matches. Since furthermore the effect of sequences decays after a
few more matches (up to 8) these observations are consistent with
the notion that the fitness does not change during a season (if
averaged over the time scale of at least a quarter season).

Of course, a further improvement of the statistical analysis could be reached if
further explanatory variables are implemented such as the possession
of the ball \cite{Hirotsu03}. It would be interesting to quantify the
increase of the predictive power in analogy to the analysis of this work;
see, e.g., Tab.I.

Whereas some of our results were expected, we had to revise some of
our own intuitive views on how professional soccer works. Using
objective statistical methods and appropriate concepts, mostly taken
from typical physics applications, a view beyond the common
knowledge became possible. Probably, for a typical soccer fan also
this statistical analysis will not change the belief that, e.g.,
his/her support will give the team the necessary impetus to the next
goal and finally to a specific home fitness. Thus, there may exist a
natural, maybe even fortunate, tendency to ignore some objective
facts about professional soccer. We hope, however, that the present
analysis may be of relevance to those who like to see the systematic
patterns behind a sports like soccer. Naturally, all concepts
discussed in this work can be extended to different types of sports.
Furthermore an extension to single-match properties as well as a
correlation with economic factors is planned for the future.

We would like to thank S.F. Hopp, C. M\"uller and W.
Krawtschunowski for the help in the initial phase of this project
as well as B. Strauss, M. Tolan, M. Trede and G. Schewe for
interesting and helpful discussions. Furthermore we would like
to thank H. Heuer for bringing the work of Atkinson to our attention.

\section{Appendix I}

Here we consider a simple model which further rationalizes the
statement that observables with larger Pearson correlation
coefficients (correlation between first and second half of season)
are better measures for the fitness of a team. This holds
independent of whether or not the true fitness changes during a
season or remains constant. We assume that the true fitness of a
team $i$ at time $j$ ($j$ may either reflect a single match or, e.g.,
the average fitness during the $j$-th half of the season) can be
captured by a single number $\mu_{i,j}$. Evidently, the true
fitness $\mu_{i,j}$ of team $i$ is not exactly known. The variance
of the fitness $\sigma_\mu^2$ is assumed to be time independent,
which just reflects stationarity.

In the experiment (here: soccer match) one observes the outcome
$x_{i,j}$ which may, e.g., correspond to the goal difference or
the number of points of team $i$ at time $j$. We assume a
Markovian process, i.e. the outcome at time $j$ is not influenced
by the outcome in previous matches. Naturally $x_{i,j}$ is
positively correlated with $\mu_{i,j}$. Without loss of generality
we assume that the $\langle \mu_{i,j} \rangle_i = \langle x_{i,j}
\rangle_i = 0$. The index $i$ reflects the fact that the averaging
is over all teams. For reasons of simplicity we assume a
linear relation between $x_{i,j}$ and $\mu_{i,j}$, namely
\begin{equation}
\label{app1}
x_{i,j} = a (\mu_{i,j} + \xi).
\end{equation}
Here $a > 0$ is a fixed real number and $\xi$ some noise,
characterized by its variance $\sigma_\xi^2$. The noise reflects the
fact that the outcome of a soccer match is not fully determined by
the fitness of the teams but also includes random elements. This relation
expresses the fact that a team with a better fitness will on average also
perform better during its matches.

The key idea in the present context is to use the outcome of matches to {\it estimate}
the team fitness.
The degree of correlation between  $x_{i,j}$ and $\mu_{i,j}$
is captured by the correlation coefficient
\begin{equation}
\label{app2}
c_{x_j,\mu_j} = \frac{\langle x_{i,j} \mu_{i,j} \rangle_i}{\sigma_x
\sigma_\mu}.
\end{equation}
A large value of $c_{x_j,\mu_j}$ implies that the estimation of $\mu_{i,j}$, based on
knowledge of $x_{i,j}$ works quite well. Thus, one may want to search for observables
$x_{i,j}$ with large values of  $c_{x_j,\mu_j}$. Unfortunately, since
$\mu_{i,j}$ cannot be measured $c_{x_j,\mu_j}$ is not directly
accessible from the experiment. The theoretical expectation reads (see Eqs. \ref{app1} and \ref{app2})
\begin{equation}
\label{cxm} c_{x_j,\mu_j} = \frac{\sigma_\mu}{\sqrt{\sigma_\mu^2
+ \sigma_\xi^2}}.
\end{equation}

For a closer relation to the general experimental situation one has to take into
account that the team fitness may somewhat change with time. This can be generally
captured by the correlation factor
\begin{equation}
c_{\mu_j,\mu_{j+1}} = \frac{\langle \mu_{i,j+1} \mu_{i,j} \rangle}{\sigma_\mu^2}.
\end{equation}

Experimentally accessible is the correlation of $x_{i,j}$ for two
subsequent time points $j$ and $j+1$ . A short and straightforward
calculation yields (using Eq.\ref{cxm})
\begin{equation}
c_{x_j,x_{j+1}} = c_{\mu_j,\mu_{j+1}}[c_{x_j,\mu_j}]^2.
\end{equation}
This result shows that {\it independent} of the possible decorrelation of
the true fitness $\mu$ observables $x$ with a larger correlation coefficient $c_{x_j,x_{j+1}}$
display larger $c_{x_j,\mu_j}$, i.e. form a better measure for the true
 fitness $\mu$. This is the line of reasoning used to
identify $\Delta G$ as a better fitness measure than the number of points independent of whether
or not $\Delta G$ changes during a season.

To go beyond this key statement we specify the loss of correlation of the
true fitness via the simple linear ansatz
\begin{equation}
\mu_{i,j+1} = b \mu_{i,j} + \epsilon.
\end{equation}
Here the noise term is characterized by the variance $\sigma_\epsilon^2$.
For reasons of simplicity we assume that the random-walk type
dynamics is identical for all teams. Stationarity is guaranteed
exactly if
\begin{equation}
\sigma_\epsilon^2 = \sigma_\mu^2(1 - b^2).
\end{equation}
Constant fitness naturally corresponds to $b=1$ and
$\sigma_\epsilon = 0$.
Of particular interest for the present work is the average of
$x_{i,j}$ over $N$ times (e.g. $N$ matches if $j$ counts the
matches). Here we define
\begin{equation}
X_{i}(N) = \frac{\sum_{j=1}^N x_{i,j}}{N}.
\end{equation}
The variance of this average, denoted $\sigma_{X(N)}$ can be
calculated in a straightforward manner. The result reads
\begin{equation}
\label{X_exact} \sigma_{X(N)}^2 = \frac{a^2\sigma_\mu^2}{N^2}
\left [ N + 2b\frac{N-1-Nb+b^N}{(1-b)^2}\right ] + \frac{a^2 \sigma_\xi^2}{N}.
\end{equation}
For $b=1$ one obtains $\sigma_{X(N)}^2 = a^2 \sigma_\mu^2 +
a^2\sigma_\xi^2/N$. Thus, in case of constant team fitness one
gets a linear behavior in the $1/N$ representation and the limit
value just corresponds to the variance of the team fitness (apart
from the trivial constant $a$). This implies that by extrapolation
one can get important information about the underlying statistics,
as described by the true team fitness $\mu_{i,j}$. This just
reflects the fact that for sufficient averaging the noise effects
become irrelevant. For $b < 1 $, however, one has a crossover from
that behavior to $\sigma_{X(N)}^2 = a^2
\sigma_\mu^2[(1+b)/(1-b)]/N + a^2\sigma_\xi^2/N$ for large $N$,
thus approaching zero for large $N$. Since $\sigma^2_{\Delta
G}(N)$ did not show any bending we have concluded in the main text
that the data do not indicate a decorrelation of the fitness
within a single season.

\section{Appendix II}

\begin{figure}
\includegraphics[width=7cm]{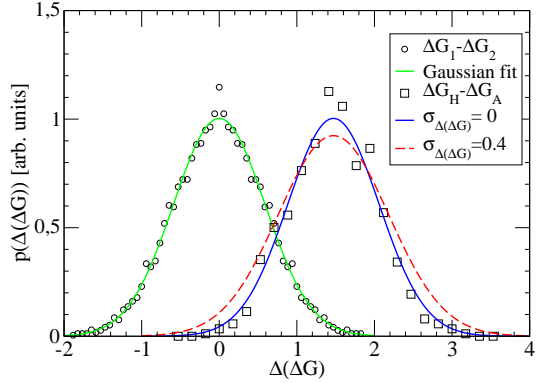}
\caption{\label{heimstaerke} Analysis of the home fitness. The
squares correspond to the actual distribution of $\Delta (\Delta
G)$. This curve is compared with the estimation for
$\sigma_{\Delta (\Delta G)} = 0$ and $\sigma_{\Delta (\Delta G) }
= 0.4$. For more details see text.}
\end{figure}

Here we discuss in more detail the distribution of $\Delta(\Delta
G)(N=17)$ shown in Fig.\ref{heimstaerke}. Of course, it has
a finite width due to statistical effects. Our goal is to compare
this distribution with a second distribution which is generated
under the assumption that no specific home fitness exists. For
this purpose we have defined, for each team in a given season, the
random variable $\Delta G_1 - \Delta G_2$. Here the first term
contains the average of the goal differences of some 17 matches
and the second term the average over the remaining 17 matches. The
34 matches were attributed to both terms such that the number of
home matches of the first term is 9 (or 8) and that of the second
team is 8 (or 9), respectively. Then we have generated the
distribution of $\Delta G_1 - \Delta G_2$. In order to get rid of
the residual home effect (9 vs. 8) we have shifted this curve so
that the average value is 0. This procedure has been repeated for
many different mappings of this kind and for all seasons. The
resulting curve is also shown in Fig.\ref{heimstaerke}. It
reflects the statistical width of $\Delta (\Delta G)$ after a
season if no home advantage were present. It can be very well
described by a Gaussian.  When shifting this distribution by the
value of the average home advantage one obtains an estimate of the
distribution of $\Delta (\Delta G)$ for $\sigma^2_{\Delta (\Delta
G)}= 0$. To be consistent with this procedure we have generated the
distribution of $\Delta (\Delta G)(N=17)$ in an analogous way. We
have calculated this distribution for every individual season and
shifted each curve so that the mean agrees with the overall mean.
 In this way we have removed a possible broadening of this
curve due to the year-to-year fluctuations of the general home
advantage.

In agreement with the discussion of Fig.\ref{heimst_n} one
observes a good agreement with the actual distribution of $\Delta
(\Delta G)$. By convolution of this distribution with a Gaussian
with variance $\sigma^2_{\Delta (\Delta G)}$ one can get
information about the sensitivity of this analysis. Choosing, e.g.,
 $\sigma_{\Delta (\Delta G)} = 0.4$, one can clearly see
that this choice is not compatible with the actual distribution of
$\Delta (\Delta G)$. Thus, if at all, the residual home fitness
can be described by a value of $\sigma_{\Delta (\Delta G)}$
significantly smaller than 0.4. In the main text we have derived
an upper limit of 0.12.


\end{document}